\documentclass[]{aa} 
\usepackage{graphicx}
\usepackage{txfonts}

\begin{document}
\title{Forbush decreases and turbulence levels at CME fronts}
\author{P. Subramanian\inst{1}, H. M. Antia\inst{2}, S. R. Dugad\inst{2},
U. D. Goswami\inst{2}, S. K. Gupta\inst{2}, Y. Hayashi\inst{3}, N. Ito\inst{3},
S. Kawakami\inst{3}, H. Kojima\inst{4}, P. K. Mohanty\inst{2}, P. K.
Nayak\inst{2}, T. Nonaka\inst{3}, A. Oshima\inst{3}, K. Sivaprasad\inst{2},
H. Tanaka\inst{2}, S. C. Tonwar\inst{2}\\{\large (The GRAPES-3 Collaboration)}}
\authorrunning{Subramanian et al.}

\institute{Indian Institute of Science Education and Research, Sai Trinity Building, Pashan, Pune 411021, India \and
Tata Institute of Fundamental Research, Homi Bhabha Road,
Mumbai 400005, India \and
Graduate School of Science, Osaka City University, Osaka 558-8585, Japan \and
Nagoya Women's University, Nagoya 467-8610, Japan}

   \date{}

  \abstract 
  {} 
{We seek to estimate the average level of MHD turbulence near coronal
mass ejection (CME) fronts as they propagate from the Sun to the Earth.}
{We examined the cosmic ray data from the GRAPES-3 tracking muon telescope
at Ooty, together with the data from other sources for three closely observed
Forbush decrease events. Each of these event is associated with
frontside halo coronal mass ejections (CMEs) and near-Earth magnetic
clouds. The associated Forbush decreases are therefore expected to
have significant contributions from the cosmic-ray depressions
inside the CMEs/ejecta. In each case, we estimate the magnitude of
the Forbush decrease using a simple model for the diffusion of high-energy
protons through the largely closed field lines enclosing the
CME as it expands and propagates from the Sun to the Earth. The
diffusion of high-energy protons is inhibited by the smooth,
large-scale magnetic field enclosing the CME and aided by the
turbulent fluctuations near the CME front. We use estimates of the
cross-field diffusion coefficient $D_{\perp}$ derived from the published
results of extensive Monte Carlo simulations of cosmic rays
propagating through turbulent magnetic fields. We then compare our
estimates with the magnitudes of the observed Forbush decreases.}
{Our method helps constrain the ratio of energy density in the
turbulent magnetic fields to that in the mean magnetic fields near the
CME fronts.  This ratio is found to be $\sim$ 2\% for the 2001 April 11
Forbush decrease event, $\sim$ 6\% for the 2003 November 20 Forbush
decrease event and $\sim$ 249\% for the much more energetic event of
2003 October 29.}  {}


\maketitle
\section{Introduction}
Forbush decreases are short-term depressions in the cosmic ray
flux reaching the Earth, and they are caused by the effects of the
interplanetary counterparts of coronal mass ejections (CMEs) from
the Sun (and the shocks they drive) and also the co-rotating
interaction regions originating from the Sun. They have been
studied closely since their initial discovery in the 1930s (see, e.g.,
comprehensive observational reviews by Cane 2000; Venkatesan \&
Badruddin 1990). With the recent upsurge of interest in space
weather effects due to solar transients, the complementary
information provided by the cosmic ray signatures of these effects
has assumed increased significance.

In this work, we have examined the data from the GRAPES-3 tracking
muon telescope at Ooty for three well-observed events. Earlier, we
had analysed the Forbush decrease data from this experiment during
the period 2001--2004 and searched for events that can be associated
with a near-Earth magnetic cloud and a corresponding halo CME from the
Sun. We start from lists of magnetic cloud events given on the WIND
webpage, http://lepmfi.gsfc.nasa.gov/mfi/mag\_cloud\_pub1.html, Lynch
et al.~(2003) and Huttunen et al.~(2005). We shortlist Forbush
decrease events that occurred in a reasonable time window around the
time of entry of the magnetic clouds. Of these shortlisted events, we
have  selected the three events shown in Fig.~\ref{fignm}. Two
of these events (2001 April 11 and 2003 October 29) have
especially clean decrease profiles. Even though the decrease profile
for the third event on 2003 November 20 is not nearly as clear-cut,
we still selected it, because of its association with a very closely
observed magnetic cloud, and also because it illustrates the diversity
of the events observed with the GRAPES-3 experiment. We estimate the
contribution of the depressed cosmic ray density inside the CME to the
total Forbush decrease seen in these datasets, and use it to derive
conclusions regarding the level of turbulence near the CME front.

The events discussed in this work were very closely observed, and
they have their origins in full halo CMEs that originated close to
the centre of the solar disc. This means that the corresponding
interplanetary CMEs (ICMEs) would have been intercepted at the Earth
as ejecta/magnetic clouds, and the observed cosmic ray depressions
corresponding to these events would have had contributions from the
shock ahead of the ICME, as well as from the ejecta/magnetic cloud
itself (Cane et al.~1994; Cane et al.~1995; Cane \& Richardson 1997).
Theoretical treatments of Forbush decreases model the effect as
arising due to a general propagating region of enhanced
turbulence/scattering and decreased diffusion (e.g., Nishida 1983; le
Roux \& Potgieter 1991) and do not distinguish between the shock and
the ejecta, or implicitly assume that the decrease is only due to the
shock (e.g., Chih \& Lee 1986). It is fairly well known that magnetic
clouds are well-correlated with Forbush decreases (e.g., Zhang \&
Burlaga 1988; Badruddin et al.~1986;  Badruddin et al.~1991;
Venkatesan \& Badruddin 1990; Sanderson et al.~1990; see, however,
Lockwood et al.~1991 for the opposite viewpoint). 

The relationship between the decrease due to the shock and the one due
to the magnetic cloud has been investigated (Cane et al.~1995, 1997;
Wibberenz et al.~1997, 1998), and there is some
evidence that they can contribute in roughly equal proportions, to the
overall magnitude of the decrease, although the individual time
profiles of the decrease due to these two effects can be quite
different (Wibberenz et al~1997). It is worth mentioning that the
contributions of the shock and the magnetic cloud to the overall
decrease can be discerned relatively easily from the spacecraft
observations (e.g., Cane et al.~1995, 1997; Cane~2000), since the
temporal boundaries of the magnetic cloud can be readily ascertained
by using the data from the same source. These data typically exhibit
two steps in the decrease phase, one corresponding to the shock
passage and the second to the entry of the magnetic
cloud. The two-step signature, however, need not be a necessary
condition for the Forbush decreases arising from the combination of a
magnetic cloud and the associated shock. Even for events where there
is clear evidence that both the shock and the following
magnetic cloud have intercepted the Earth (as is the case with the three
events presented in this paper), the precise decrease profile will
depend upon the standoff distance between the magnetic cloud and the
shock and on the energies of the protons detected. If the protons are
energetic enough that their mean free path is comparable to the
magnetic cloud-shock standoff distance, the distinct identities of the
steps corresponding to the shock and the cloud may not be clear. We
attribute 50\% of the total decrease to the magnetic cloud
($\alpha = 0.5$, see \S~3), following the general logic outlined by
Wibberenz et al. (1997). We discuss the sensitivity of our final
results to variations in $\alpha$ in \S~6. In our quantitative work,
we estimate the magnitude of the Forbush decrease using filtered data
(see \S~2.1 for a  description of the filtering method) averaged over
1 hr.

The high muon counting rate measured by the GRAPES-3 experiment
results in extremely small statistical errors, allowing small
changes in the intensity of the cosmic ray flux to be measured
with high precision. Thus a small drop ($\sim$0.2\%) in the
cosmic ray flux, during a Forbush decrease event can be reliably
detected. This is possible even in the presence of the diurnal
anisotropy of much larger magnitude ($\sim$1.0\%), through a
suitable filtering technique described subsequently.

Before we proceed further, it is worth discussing the terminology we
use. The term CME is generally used to denote the blob of plasma ejected
from the solar corona as viewed by coronographs near the Sun
(typically from 1 to 30 $R_{\odot}$). The CME can be thought of as a
largely closed, magnetic flux rope-like magnetic structure (see,
however, Bothmer et al.~1996, 1997 for evidence of CMEs possessing
closed, as well as some open field lines) whose cross-section expands
as it travels through the heliosphere. There are practically no high-energy
cosmic rays inside it when it starts out from near the Sun, and
the ambient cosmic rays diffuse into it via cross-field diffusion
across the closed magnetic field lines as it travels through the
heliosphere. The term ``ICME'' is a broad one used to denote the
interplanetary counterpart of a CME (see, e.g., Forsyth et al.~2006;
Wimmer-Schweingruber et al.~2006 and references therein), while the
term ``magnetic cloud'' is reserved for ICMEs detected near the Earth,
possessing certain well-defined criteria such as plasma temperature
depressions and smooth magnetic field rotations (see, e.g., Burlaga et
al.~1981, Bothmer \& Schwenn 1998). Other CME/ICME related ejecta
detected near the Earth are merely called ``ejecta''.

The rest of this paper is organised as follows, We briefly describe
the GRAPES-3 experiment in \S~2. We describe the model we used to
calculate the magnitude of the expected Forbush decreases in \S~3. The
three Forbush decrease events we analysed are described in \S~4,
5, and 6. We describe our results in \S~7 and summarise them in \S~8.

\section{GRAPES-3 Experimental system}

\begin{figure}
\begin{center}
\includegraphics*[width=\columnwidth,angle=0,clip]{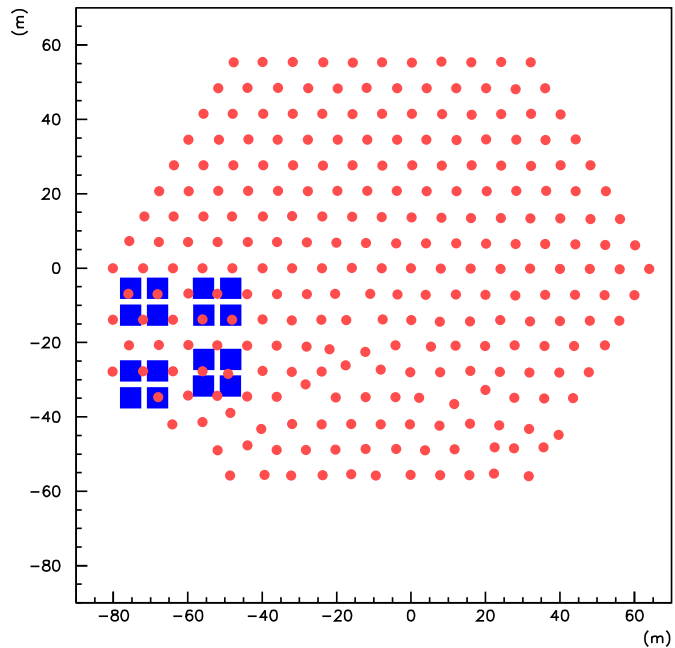}
\vskip -0.1in
\caption{\label{fig1} A schematic layout the GRAPES-3 air shower
         array shown here with 217 detectors (circles). Each
	 of the 16 squares represent a 35 m$^2$ area tracking muon
	 detector with $E_{\mu}\ge1$ GeV used in the present work.}
\end{center}
\end{figure}

The GRAPES-3 experiment is located at Ooty ($11.4^{\circ}$N
latitude, $76.7^{\circ}$E longitude, and 2200 m altitude), in
southern India. The GRAPES-3 air shower experiment was designed
to have a compact configuration of a conventional
type array, with a separation of only 8 m between the adjacent
detectors, which are deployed in a symmetric hexagonal geometry.
A schematic layout of the GRAPES-3 array is shown in Fig.~\ref{fig1}.
The observations were started in 2000 with 217 detectors, located
within the inner 8 rings, see Fig.~\ref{fig1} (Gupta et al.~2005).

A very large-area tracking muon telescope operating as a part of
the GRAPES-3 experiment (Gupta et al.~2005; Hayashi et al.~2005),
is a unique instrument, to search for the high-energy protons
emitted during the active phase of a solar flare or a coronal mass
ejection (CME). The muon telescope is capable of providing a
high-statistics, directional study of muons. The GRAPES-3 muon
telescope covers an area of 560 m$^2$, consisting of a total 16
modules, each 35 m$^2$ in area. These modules are located close to
each other as shown in Fig.~\ref{fig1}. A cluster of four 35 m$^{2}$
area neighbouring modules, located inside a common hall, constitutes
one supermodule with a total area of 140 m$^{2}$. The energy threshold
of the telescope is 1 GeV for the muons arriving along the vertical
direction. The cutoff rigidity due to the magnetic field of the Earth
at Ooty is 17 GV in the vertical direction and varies from 12 to 42
GV across the field of view of the muon telescope as shown later in
Fig.~\ref{fig6}.

\begin{figure}
\begin{center}
\includegraphics[width=\columnwidth]{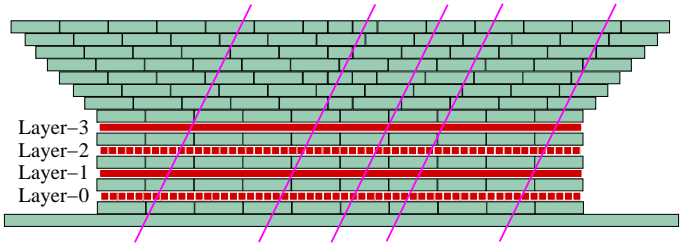}%
\caption{\label{fig2} A schematic display of the 4-layer tracking 
         muon telescope module with 58 PRCs per layer. The four
         layers of the PRCs labelled Layer-0, Layer-1, etc. are
	 embedded in concrete blocks. Inclined lines represent a
         set of parallel muon tracks.}
\end{center}
\end{figure}

The basic detector element of the muon telescope is a rugged
proportional counter (PRC) made from a 600 cm long steel pipe
with 2.3 mm wall thickness and a square cross-sectional
area of $10\times10$ cm$^2$. A muon telescope module with a
sensitive area of 35 m$^2$ consists of a total of 232 PRCs
arranged in four layers of 58 PRCs each, with alternate layers
placed in orthogonal directions. Two successive layers of the
PRCs are separated by a 15 cm thick concrete layer, consisting
of 60$\times$60$\times$15 cm$^3$ blocks as shown in Fig.~\ref{fig2}.
The four-layer PRC configuration of the muon modules allows
a 3-D reconstruction of the muon track direction to an accuracy
of $\sim6^{\circ}$. It is worth noting that the accuracy
gradually increases with increasing zenith angle due to the
greater separation of the triggered PRCs.

To achieve an energy threshold of 1 GeV for vertical muons, an
absorber of total thickness $\sim550$ g cm$^{-2}$ in the form
of concrete blocks is employed. This was done by placing a total
of 15 layers of concrete blocks above Layer-1, as shown
schematically in Fig.~\ref{fig2}. A unique feature of the GRAPES-3
muon module is the robust structure of the PRCs, which supports the
huge load of 2.4 m thick concrete in a self-supporting manner. The
concrete blocks are arranged in the shape of an inverted pyramid
to shield the PRCs, with coverage up to $45^\circ$ around the
vertical direction for the incident muons. The threshold energy
changes to $\sec\theta$ GeV for the muons incident at a zenith
angle of $\theta$. The cross section of a muon telescope module is
shown schematically in Fig.~\ref{fig2}. A cluster of four such
modules, separated by a horizontal distance of 130 cm at the base
constitutes one supermodule. The GRAPES-3 muon telescope contains
a total of four supermodules (Hayashi et al.~2005).

Because of the sensitivity of the PRCs to low-energy $\gamma$
rays, from the radioactivity present in the concrete absorber,
individual PRCs display sizable counting rates of $\sim200$ Hz.
An output is generated, if any one of the 58 PRCs produces a
signal. A logical `OR' of outputs from all 58 PRCs in a layer
is generated, after suitable amplification and shaping to form
the layer OR output. A coincidence of the four OR outputs from
the four layers in a module is used to generate the 4-layer
trigger. Despite high counting rates of individual PRCs, the
4-layer coincidence trigger is relatively free of the
contribution from the background radioactivity and it is caused
only by the passage of a muon. The observed 4-layer muon counting
rate of $\sim3200$ Hz per module yields a total counting rate
$\sim3\times10^6$ min$^{-1}$ for all 16 modules. This high
rate permits small changes of $\lesssim0.1\%$ in the muon flux,
to be detected on a timescale of $\sim5$ min, after appropriate
correction to the variation in the atmospheric pressure with time.

\begin{figure}
\begin{center}
\includegraphics[height=\columnwidth,angle=270]{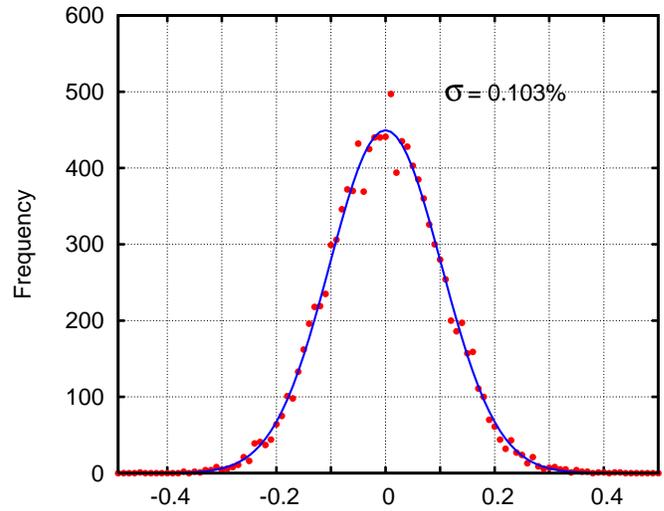}%
\caption{\label{fig3} Distribution of the relative variation
         in \%, for the 4-layer muon counting rate for the 35
	 m$^{2}$ area modules for time intervals of 6 minutes.
	 Also shown is a Gaussian fit to the data. The standard
	 deviation of the fit is 0.103\% as against an expected
	 value of 0.095\% based on statistics.}
\end{center}
\end{figure}

Most of the detected muons are generated by $\gtrsim20$ GeV galactic
cosmic rays, and form a stable and dominant background to the variation
in their flux produced by the CME/solar flare. The muon data is
grouped online every 10 sec, into solid-angle bins of $\sim0.05$ sr,
consistent with the angular resolution of the muon telescope as
described below. Since $>1$ GeV muons are secondaries produced by the
primary protons of energy $\gtrsim20$ GeV in the atmosphere, therefore
these observations can be used to probe the effect of the Sun on
cosmic rays.

The data recorded during 2003 October, from 15 out of 16 working
modules were used in the statistical analysis described below. The
muon rate for each 35 m$^{2}$ area module was recorded for 770
time intervals, each 6 minutes in duration spread over a total of
about 3.5 days. During this period of 3.5 days, the data could not
be recorded for 70 out of 840 intervals, due to the failure of the
recording system in the first supermodule. Next, the mean muon rate
of the module for the entire duration was computed. Then the percent
deviation from the mean module rate was calculated, separately for
each module. Finally, the rms spread in the percent deviation called
`relative variation' was calculated for each module for all 6-minute
intervals. In Fig.~\ref{fig3}, the distribution of the relative
variation in the 4-layer muon rate in percent, as recorded by the
muon modules is shown. Also shown in Fig.~\ref{fig3}, is a Gaussian
fit to this distribution, which yields a standard deviation of
0.103\%. This value may be compared with an expectation of 0.095\%,
based on a 6-minute muon statistic of $1.1\times10^6$ per module.
This tiny statistical error in muon rate allows a high-precision
study of various solar phenomena to be carried out
(Nonaka et al.~2006).

\begin{figure}
\begin{center}
\includegraphics[width=\columnwidth]{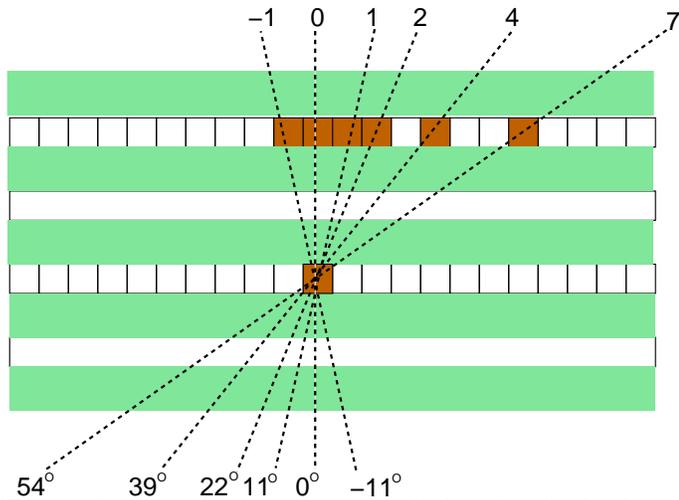}%
\vskip -0.1in
\caption{\label{fig4} A schematic view of muon arrival angle
         selection based on the PRC triggered in the lower and 
         15 PRCs in the upper layer. The triggered PRCs are
	 shown as filled squares.}
\end{center}
\end{figure}

\begin{figure}
\begin{center}
\includegraphics[width=\columnwidth]{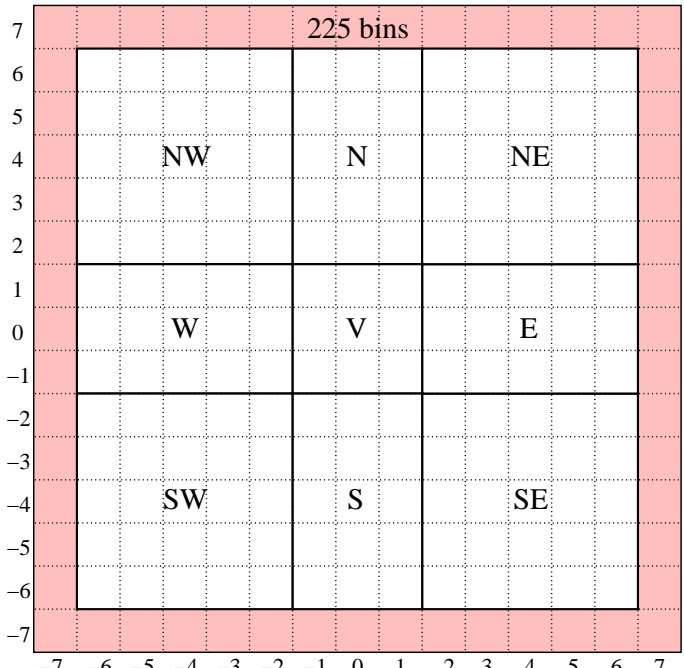}%
\vskip -0.1in
\caption{\label{fig5} A schematic of the nine muon arrival
         direction bins; $3\times3$ vertical bin V, and four
	 $3\times5$ central bins N, E, W, S, and four
	 $5\times5$ outer bins NE, SE, SW, NW.}
\end{center}
\end{figure}

\begin{figure}[!htb]
\begin{center}
\includegraphics[width=\columnwidth]{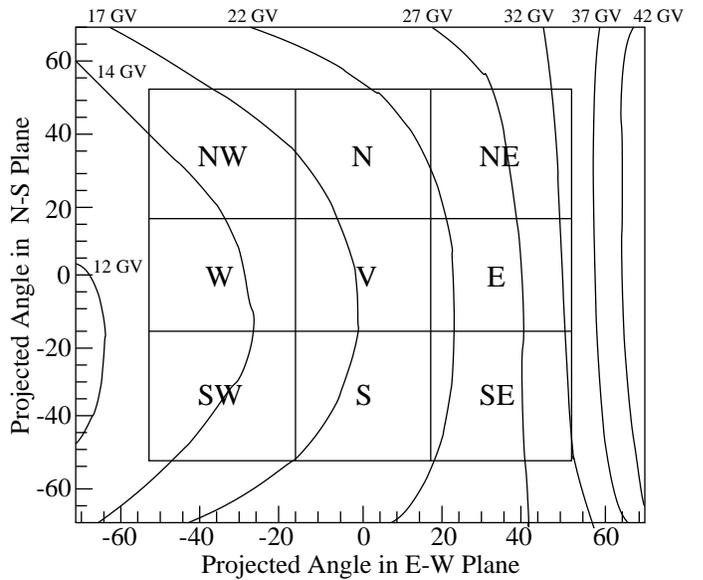}%
\vskip -0.1in
\caption{\label{fig6} The 9 coarse solid-angle bins are shown
         along with the contours of constant geomagnetic cutoff
	 rigidity in the field of view (FOV). Cutoff rigidity
	 varies from 12 to 42 GV in the FOV of GRAPES-3.}
\end{center}
\end{figure}
	 
\subsection{Data analysis}

In all four supermodules, the direction of muons is recorded into
225 solid-angle bins. This was done by using a dedicated
direction-sensitive trigger with an independent data acquisition
system for each of the four supermodules. As shown in Fig.~\ref{fig4},
the muon angle is determined for each PRC in the lower layer and
binned into 15 angular bins based on the specific location of the
PRC triggered in the upper layer from among the 15 PRCs, one directly
above (central PRC) and 7 each on either side of the central PRC.
This angular binning is carried out in each of the two orthogonal
projections (XZ and YZ; Z is vertical direction), thereby generating
a 2-dimensional $15\times15=225$ solid-angle map of muon directions.
The contents of the 225 solid-angle bins are recorded, once every 10
sec, thus providing a continuous monitoring of the directional flux
of muons in the sky.

The variation in the muon rate may be studied in any of the 225 solid
angle bins. However, it is expected that the influence of  a solar
flare and/or CME would be spread over several bins. This directional
spread could arise from the influence of the terrestrial, solar, and
interplanetary magnetic fields, etc. Therefore, the detected muons have
been regrouped into $3\times3 = 9$ coarse solid-angle bins, as shown
schematically in Fig.~\ref{fig5}. This regrouping of the data was done
by combining either a set of $3\times5$ or $5\times5$ fine solid-angle
bins. The exception being the vertical direction where $3\times3$ bins
have been combined. This choice of angular segmentation was dictated
by the fact that the muon flux is comparatively larger for the near
central directions (N, E, W, S) than for the outer directions (NE, SE,
NW, SW). This choice results in a relatively similar solid-angle
coverage for the nine coarse bins. Thus the solid-angle of acceptance
includes only $13\times13 = 169$ out of original 225 bins, restricting
the maximum zenith angle to $50^{\circ}$. This still exceeds the
shielding coverage of $45^{\circ}$, for the PRCs at the outer edge,
but such events constitute $<$1\% of the data. This regrouping also
results in muon statistics for various bins that are not too dissimilar.

At the energies ($\gtrsim20$ GeV) of interest here, the propagation
of charged particles near the Earth ($< 20 R_E$, where $R_E$ is
the radius of the Earth) is strongly influenced by the geomagnetic
field. The access by a charged particle to a given geographical
location depends on the momentum per unit charge of the particle
called rigidity. The threshold value of the rigidity is termed
``geomagnetic cutoff rigidity'', which depends on the geographical
location on the Earth and the direction of the arriving particle. The
geomagnetic cutoff rigidity can be calculated using a detailed model
of the geomagnetic field (Cooke et al.~1991). The geomagnetic cutoff
rigidity for the field of view (FOV) of the GRAPES-3 muon telescope
varies significantly for the nine coarse, solid-angle bins. We first
calculate the cutoff rigidity for the centre of each of the 169 fine
solid-angle bins, which constitute the 9 elements of the FOV, using
the IGRF2000 geomagnetic field model. Subsequently, a weighted mean
of the cutoff rigidities of the fine bins constituting a coarse bin 
is calculated for each of the 9 coarse bins. These weights are the
muon counting rates for a given fine bin. Here, it needs to be
emphasised that the knowledge of the geomagnetic field at any
particular moment in time is imperfect, and it is virtually
impossible to determine the cutoff to a high degree of accuracy.
However, the calculated values of the geomagnetic cutoff represent
a very useful approximation to the true values at the time of the
observations. In Fig.~\ref{fig6} the contours of constant
geomagnetic cutoff rigidity in the FOV are superimposed over a
schematic of the 9 solid-angle bins of muon arrival directions. The
geomagnetic cutoff rigidity varies from 12 GV in the west to 42 GV
in the east, within the FOV of the GRAPES-3 tracking muon telescope.

\begin{figure}[!htb]
\begin{center}
\includegraphics[width=\columnwidth]{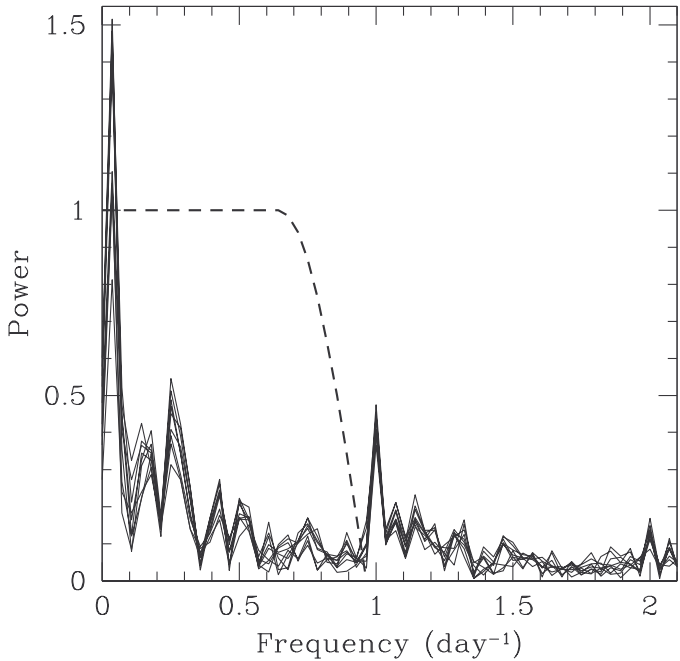}%
\vskip -0.1in
\caption{\label{fig7} The power spectrum of 28 day data
         covering the period from 26 March to 22 April 2001.
	 The solid lines show the results for all 9 directions
	 used in the study while the dashed line shows the
	 function that is used for filtering out the high-frequency 
	 components. The Fourier transform is multiplied by 
	 this function before taking the inverse Fourier transform
	 to get filtered time series.}
\end{center}
\end{figure}

In this study we used the GRAPES-3 data summed over a time interval of
one hour for each of the nine bins, which are identified as NW, N, NE,
W, V, E, SW, S, and SE. The calculated cutoff rigidities for these nine
bins are 15.5, 18.7, 24.0, 14.3, 17.2, 22.4, 14.4, 17.6, and 22.4 GV,
respectively, as seen from Fig.~\ref{fig6}. The summing over an
interval of 1 hour improves the signal-to-noise ratio, but the diurnal
variations in the muon flux are still present. To filter out these
oscillations we applied a low-pass filter, removing all frequencies
higher than 1 day$^{-1}$. In Fig.~\ref{fig7} we show the Fourier
transform of the data covering a period of 28 days that includes the
event of 2001 April 11. The peaks corresponding to diurnal variation
and its first harmonic are visible. The Fourier transform is multiplied
by the function shown by a dashed line to remove the high-frequency
components. This filter is found to be effective in removing
high-frequency oscillations, including the diurnal variations and
their harmonics. The Forbush decrease events are also clearer
in the filtered data.

Although the smoothing may tend to change the
amplitude of the decrease and possibly shift the onset time for the
Forbush decrease by a few hours in some cases. However, it is often
difficult to determine if these differences are artifacts of smoothing
or whether the unfiltered data showed different amplitude because a
diurnal oscillation happened to have the right phase, so as to enhance
or reduce the amplitude of the Forbush decrease. Some fluctuations in
muon flux could be due to Forbush decrease and associated events, but
it is unlikely that these will be periodic in nature, hence are not
likely to be affected by the filter. Since the differences in amplitude
caused by filter are not substantial, it would not affect our results
significantly. Thus in this work, we used the filtered data to study
the characteristics of Forbush decrease events, although the unfiltered
data are also shown in corresponding figures for direct comparison.
Further, we repeated the calculations using unfiltered data and find
that the final results are not significantly different. There can also
be anisotropies intrinsic to the CME/magnetic cloud itself, arising
from a $\vec{B} \times \vec{\nabla}\,N$ drift, where $\vec{B}$ is the
interplanetary magnetic field and $\vec{\nabla}\, N$ denotes the
cosmic-ray density gradient inside the CME (Bieber \& Evenson 1998;
Munakata et al.~2003, 2005; Kuwabara et al.~2004). Such anisotropies
can potentially be ``mixed'' with the diurnal anisotropy, and it is
possible that there will still be some residual anisotropy after the
filter is applied. This is the case with one of the weaker events that
we have studied here; namely 2003 November 20.

To determine the properties of Forbush decrease, we divided the
time interval in three parts, the first part before the onset;
the second part includes the decrease, while the third part
includes the recovery phase. We fit straight lines to the first
two parts and an exponential profile for the third part. The point
of intersection of the two straight lines defines the onset time
for the Forbush decrease, while the intersection of the last two
fits determines the minimum of the profile. The difference in the
flux between these two points determines the magnitude $M$ of the
decrease (\S~3), which is measured in term of percentage change in
steady flux. The fits for all events used in this work are shown
in the respective sections.

\begin{figure*}
\begin{center}
\includegraphics*[width=\textwidth,angle=0,clip]{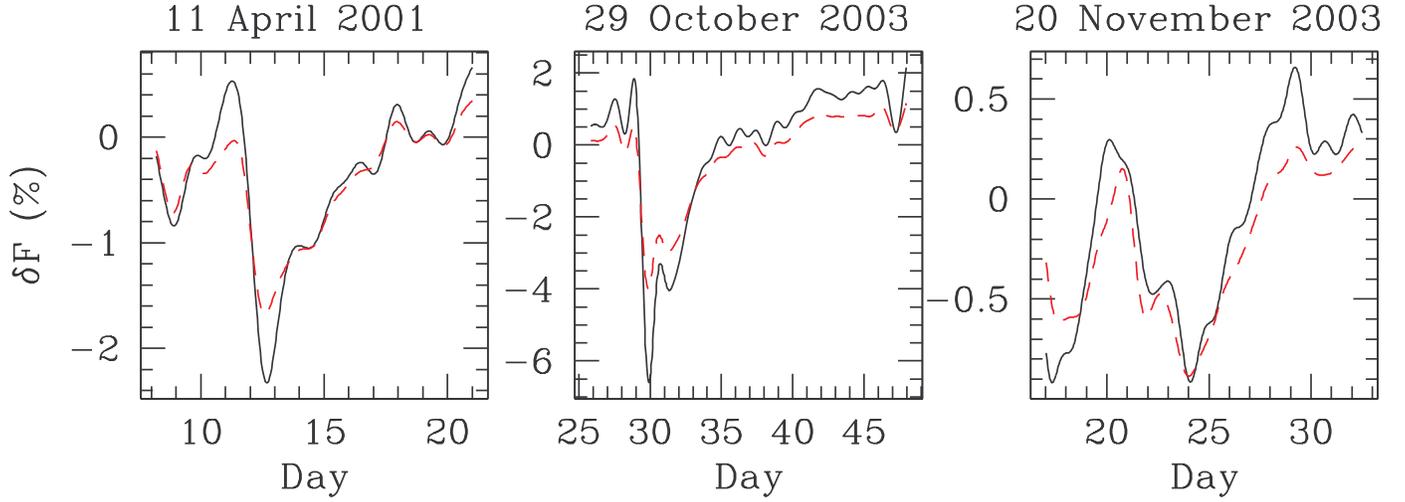}
\vskip -0.1in
\caption{\label{fignm} Data from Tibet neutron monitor (dashed
         lines) is compared with GRAPES-3 data (solid line) for
	 vertical direction for the 3 events studied in this work.
	 Both the time series are filtered by applying a low pass
	 filter to remove high frequency components. In order to
	 fit in the same scale the neutron monitor data is scaled
	 down by a factor of 3.}
\end{center}
\end{figure*}

In the subsequent analysis we discuss in detail three Forbush
decrease events, namely those on 2001 April 11, 2003 October 29,
and 2003 November 20, which were associated with CME/magnetic
clouds. The GRAPES-3 data is recorded along nine directions as
described earlier. However, in Fig.~\ref{fignm} the data for only
the vertical direction is shown for these three events along with
measurements from the Tibet neutron monitor at Yangbajing
(Muraki et al.~2007) as obtained from
http://cr0.izmiran.rssi.ru/tibt/main.htm. The Tibet neutron
monitor has a wide solid-angle of acceptance with the dominant
contribution coming from the vertical direction; for that reason,
we selected the GRAPES-3 data along the vertical direction for a
comparison as shown in Fig.~\ref{fignm}. The Tibet neutron monitor
was chosen, as Yangbajing ($90.5^{\circ}$E) is located at a similar
longitude as Ooty ($76.7^{\circ}$E), and the geomagnetic cutoff
rigidities are also not too dissimilar (Ooty=17 GV, and Yangbajing=14
GV). The neutron monitor data was scaled down by a factor of 3
to fit the same scale. Here, it needs to be emphasised that both
sets of data were subjected to the same low pass filter to remove
frequencies higher than 1 day$^{-1}$. The detailed features in the
Tibet neutron monitor data follow the GRAPES-3 muon data rather closely 
for all three Forbush decrease events as seen from Fig.~\ref{fignm}.

\section{Proton diffusion into CME/magnetic cloud}
We are concerned here with the component of the Forbush decrease
produced by the difference between the proton density inside and
outside the magnetic cloud that intercepts the Earth. To do this,
we obtain an estimate of the proton density inside the CME/magnetic
cloud produced by the cumulative effect of protons diffusing into
the closed magnetic structure of the CME/magnetic cloud as it
propagates towards the Earth. We use cross-field diffusion
coefficients derived from recent numerical treatments of cosmic ray
propagation through turbulent magnetic fields.

Before doing so, it is worth taking a brief look at the typical
proton gyro-radius in relation to the overall size of the ejecta to
see how far such a picture involving cross-field diffusion would be
valid. This is especially pertinent here, since the protons detected
by the GRAPES-3 experiment are typically very energetic. The GRAPES-3
experiment can detect muons produced by protons of rigidities above
12--42 GV, where the rigidity $Rg$ is defined by
\begin{equation}
Rg \,({\rm volts}) = \frac{P\,c}{Z\,e} = 
300\,B\,({\rm gauss})\,r_{L}\,({\rm cm})\, ,
\label{eq1}
\end{equation}
where $P$ is the proton momentum, $c$ the speed of light, $Z$ the
charge state ($= 1$ for a proton), $e$ the charge of an electron, $B$
the magnetic field in Gauss, and $r_{L}$ the proton gyro-radius in
cm. The magnetic field of a typical near-Earth magnetic cloud is
$\sim 10^{-4}$~G, while its radius is $\sim 0.2$~AU, or
$\sim 3\times10^{12}$~cm
(http://lepmfi.gsfc.nasa.gov/mfi/mag\_cloud\_Avg.html). The
gyro-radius of a 30 GV proton in the field of a typical near-Earth
magnetic cloud is $\sim 10^{12}$ cm (from Eq~\ref{eq1}), or 0.3 times
the magnetic cloud radius. Estimates of the magnetic field of
a CME near the Sun are notoriously hard to come by, but the rough
lower bound of $\sim$ 0.1 G given by Bastian et al.~(2001) can be taken
as a working number. The gyro-radius of a 30 GV proton in such a
magnetic field is $\sim 3 \times 10^{8}$ cm, which is around 3 orders of
magnitude smaller than the typical size of a CME at such heights,
which is a few solar radii (a few times $10^{11}$ cm). Thus the
gyro-radius of the typical high-energy proton whose signature is
detected by the GRAPES-3 array ranges from $10^{-3}$ to 0.3 times the
macroscopic size of the structure into which it penetrates, and the
picture of cross-field diffusion is generally valid.

The flux of protons entering the CME/magnetic cloud at a given time is
\begin{equation}
F\,({\rm cm^{-2}\,s^{-1}})\, = D_{\perp}\,\frac{\partial N_{a}}{\partial r} \, ,
\label{eq2}
\end{equation}
where $F$ is in units of number per ${\rm cm}^{2}$ per sec,
$D_{\perp}$ is the cross-field diffusion coefficient, and
$N_{a}$ the ambient density of high-energy protons. This
diffusion takes place throughout the cross-section of the
expanding CME, so the total number of high-energy protons
that will have diffused into the CME after a time $T$ is
\begin{equation}
U_{i} = \int_{0}^{T} A(t)\,F(t)\,dt = \int_{0}^{T} D_{\perp}\,A(t)\,\frac{\partial N_{a}}{\partial r} dt \, ,
\label{eq3}
\end{equation}
where $A(t)$ is the cross-sectional area of the CME at time $t$. The
integration extends from the time the CME is first observed in the
LASCO field of view ($t = 0$) through the time ($t = T$) when it
arrives at the Earth as a magnetic cloud. A reasonable estimate for
$\partial N_{a}/\partial r$ would be
\begin{equation}
\frac{\partial N_{a}}{\partial r} \simeq \frac{N_{a}}{R(t)}\, ,
\label{eq4}
\end{equation}
where $R(t)$ is the CME cross-sectional radius at time $t$. Assuming the
CME to be an expanding ``flux-rope'' whose length increases with time
as its cross-sectional area expands (see
http://lepmfi.gsfc.nasa.gov/mfi/Mag\_Cloud\_Model.html for a
cartoon), we can use
\begin{equation}
A(t) = 2 \pi \, R(t)\,L(t) \, ,
\label{eq5}
\end{equation}
where $L(t)$ is the length of the flux-rope cylinder at time $t$, and can be
related to the ``height'' $H$ of the CME above the solar limb via 
\begin{equation}
L(t) = 2\,\pi\,H(t) \, .
\label{eq5a}
\end{equation}

The height above the solar limb $H$ is the most commonly measured
quantity for limb CMEs, but it is not clear how well it can be
estimated for the halo CMEs we are considering. We consider this
aspect further in \S~4 below. Using Eqs.~(\ref{eq5}), (\ref{eq5a}),
and (\ref{eq4}) in Eq.~(\ref{eq3}), we have
\begin{equation}
U_{i} = 4 \pi^{2} \, N_{a} \int_{0}^{T} D_{\perp}\,H(t)\,dt \, .
\label{eq6}
\end{equation}
The density inside the near-Earth magnetic cloud would be
\begin{equation}
N_{i} = \frac{U_{i}}{\pi\,R(T)^{2}\,L(T)} \, ,
\label{eq7}
\end{equation} 
and it can be related to the magnitude $M$ of the Forbush decrease via
\begin{equation}
\alpha \, M = \frac{N_{a} - N_{i}}{N_{a}} = \frac{\Delta N}{N_{a}} = 1 - \frac{4\,\pi\,\int_{0}^{T} D_{\perp}\,H(t)\, dt}{R^{2}(T)\,L(T)} \, ,
\label{eq9}
\end{equation}
where $L(T)$ and $R(T)$ denote the length and cross-sectional radius
at time $T$, when the ejecta has reached the Earth. The quantity
$\alpha$ denotes the fraction of the total decrease that can be
attributed to the CME/magnetic cloud. As mentioned earlier, this
fraction is typically around 50\% (Wibberenz et al.~1997), so we adopt
$\alpha = 0.5$ in our calculations. We discuss the effect of
variations in $\alpha$ on our final result in \S~7.  

\subsection{Cross-field diffusion coefficient}
We now turn our attention to the cross-field diffusion coefficient
$D_{\perp}$, which governs the diffusion of the ambient high-energy
protons into the CME across the magnetic fields that enclose it.
Protons can travel {\bf perpendicular} to the direction of the average
magnetic field as a result of normal transport of particles due to
scattering or drift, as well as the braiding/mixing of the magnetic
field itself. The extent of cross-field diffusion depends on several
parameters, such as the proton rigidity (which indicates how tightly
the proton is bound to the magnetic field) and the level of
turbulence, which can contribute to field line transport, as well
as proton scattering.

The issue of the cross-field diffusion coefficient is a subject
of intensive ongoing research, and a good way to proceed for our
purposes is to use the convenient parametrisation of the results
of extensive Monte Carlo simulations of cosmic rays propagating
through tangled magnetic fields given by Candia \& Roulet (2004).
Their results can be regarded as a superset of similar work (e.g.,
Giacalone \& Jokipii 1999; Casse et al.~2002); it contains all the
preceding results and also extends the parameter regime of the
simulations. They give the following fit for the ``parallel''
diffusion coefficient $D_{\parallel}$ (which is due to scattering
of the particles back and forth along the mean field, as the field
is subject to random turbulent fluctuations):
\begin{equation}
D_{\parallel} = c\,L_{\rm max}\,\rho\,\frac{N_{\parallel}}{\sigma^{2}}\,\sqrt{\biggl (\frac{\rho}{\rho_{\parallel}} \biggr )^{2(1 - \gamma)} + 
\biggl (\frac{\rho}{\rho_{\parallel}} \biggr )^{2}} \, ,
\label{eq17}
\end{equation}
where $c$ is the speed of light and the quantities $N_{\parallel}$,
$\gamma$ and $\rho_{\parallel}$ are constants specific to different
kinds of turbulence whose values are given in Table 1 of Candia \&
Roulet (2004). The quantity $L_{\rm max}$ is the maximum length scale
of the turbulence. Matthaeus et al.~(2005) quote a value of $L_{\rm
max} = 10^6$ km, based on multi-spacecraft measurements of solar wind
turbulence near the Earth. This is similar to values quoted by other
authors (e.g., Manoharan et al.~1994) who use very
different methods. We take the quantity $L_{\rm max}$ to be a fixed
number from near the Sun to the Earth. Candia \& Roulet (2004) call
$\rho$ the ``rigidity'', but it is somewhat different from our usual
definition. It is related to the usual rigidity $Rg$ by
\begin{equation}
\rho = \frac{r_{L}}{L_{\rm max}} = \frac{Rg}{B_{0}\,L_{\rm max}}\, ,
\label{eq18}
\end{equation}
where $B_{0}$ is the mean magnetic field. The quantity $\sigma^{2}$ is
the turbulence level, and is defined as
\begin{equation}
\sigma^{2} \equiv \frac{\langle B_{r}^{2} \rangle}{B_{0}^{2}}
\label{eq19}
\end{equation}
where $B_{r}$ is the turbulent magnetic field and the angular braces
denote an ensemble average. The cross-field diffusion coefficient
($D_{\perp}$) is related to the parallel one ($D_{\parallel}$) by
\begin{equation}
\nonumber
\frac{D_{\perp}}{D_{\parallel}} = \cases{N_{\perp}\,(\sigma^{2})^{a_{\perp}}\, , & $\rho \leq 0.2$ \cr
\noalign{\medskip}
N_{\perp}\,(\sigma^{2})^{a_{\perp}}\,\biggl (\frac{\rho}{0.2} \biggr )^{-2}\, , & $\rho > 0.2$\, .\cr } 
\label{eq20}
\end{equation} 

The quantities $N_{\perp}$ and $a_{\perp}$ are constants specific to
different kinds of turbulent spectra, and are given in Table 1 of
Candia \& Roulet (2004). For concreteness, we assume a Kolmogorov
turbulence spectrum in our calculations. Equations~(\ref{eq17}) and
(\ref{eq20}) jointly define the cross-field diffusion coefficient.
The magnetic field $B_{0}$ of the CME changes (weakens) as it
propagates from the Sun to the Earth, which means that $\rho$ varies
with time. We do not know how $B_{0}$ varies with heliocentric
distance, but we can assume that the flux is frozen into the CME as it
propagates from the Sun to the Earth. It may be argued that some of
the magnetic flux will be dissipated in driving the CME. However, on
average, even the flux detected by near-Earth magnetic clouds is
enough to account for some 74\% of what is required to drive the CME
from the Sun to the Earth (Subramanian \& Vourlidas 2007). The
frozen-flux assumption is therefore fairly accurate. This means that
the product of the magnetic field and the cross-sectional area of the
CME remains approximately constant, i.e.,
\begin{equation}
B_{0} (t) = B_{\rm MC}\,\biggl [ \frac{R(T)}{R(t)} \biggr ]^{2} \, ,
\label{eq21}
\end{equation}
where $B_{\rm MC}$ is the magnetic field inside the magnetic cloud.
The quantity $R(t)$ is related to the measured expansion velocity
$V_{\rm exp}$ and the starting radius $R_{0}$ of the halo CME by
\begin{equation}
R(t) = V_{\rm exp}\,t + R_{0} \, .
\label{eq22}
\end{equation}
We note that the measured quantities $R_{0}$ and $R(t)$ are
the lateral extents of the halo CME measured in the plane of the sky,
and are thus representative of the lateral expansion of the CME
cross-section.

\subsection{Constraining $\sigma^{2}$}
For a given Forbush decrease event, we measure the quantity $M$, and
Eqs.~(\ref{eq22}), (\ref{eq21}), (\ref{eq20}), (\ref{eq18}), and
(\ref{eq17}) jointly define the cross-field diffusion coefficient
$D_{\perp}$, which can be used in Eq.~(\ref{eq9}). Unfortunately,
there is not much simplification possible in the expression for
$D_{\perp}$; the quantity $\rho$ typically straddles regimes such that
both the terms under the square root in Eq.~(\ref{eq17}) and both the
branches of Eq.~(\ref{eq20}) need to be retained, and when $D_{\perp}$
is finally substituted into Eq.~(\ref{eq9}) the integral needs to be
evaluated numerically. As we see below, we have concrete observational
values for all the quantities used in Eq.~(\ref{eq9}) except the
turbulence level $\sigma^{2}$ (Eq~\ref{eq19}), and it is this quantity
that we seek to constrain. Owing to the different cutoff rigidities in
different bins and remaining anisotropies, the magnitude $M$ of the
Forbush decrease (for a given event) differs from bin to bin.
Consequently, we obtain slightly different estimates for $\sigma^{2}$
for a given event, which can be regarded as a scatter (Tables 1, 2,
and 3). For a given bin, we adopt a single number for the proton
rigidity, which is equal to the cutoff rigidity for that bin. For
instance, we adopt $Rg = 17$ GV for the vertical bin, and the values of
$Rg$ for the other bins correspond to the respective cutoff rigidities
listed in \S~2.1. We note that this is a simplifying assumption, for
a muon detected in a given bin can in fact be produced by protons
possessing a range of energies. We next describe the characteristics
of the three events we have selected for our study.

\section{2001 April 11 event}
\subsection{Magnetic cloud}
\begin{table*}
\caption{Derived parameters for Forbush decrease for the 2001 April 11 event}
\begin{tabular}{llllllllll}
\hline
Quantity & NW & N & NE & W & V & E & SW & S & SE\\
\hline
FD magnitude & 2.51 & 2.51 & 1.87 & 2.92 & 2.71 & 2.04 & 2.58 & 2.37 & 1.79 \\
FD onset & 15:36 & 17:31 & 19:26 & 10:05 & 11:45 & 16:19 & 06:58 & 09:22 & 12:45\\
FD end$^1$ & 15:07 & 13:55 & 12:43 & 12:57 & 12:43 & 11:45 & 09:36 & 11:02 & 11:17\\
$\sigma^{2}$ (Kolmogorov) & 0.02 & 0.02 & 0.02 & 0.02 & 0.02 & 0.02 & 0.02 & 0.02 & 0.02\\
\hline
\noalign{\smallskip}
\noalign{$^1$ The FD end times refer to 2001 April 12.}
\end{tabular}
\end{table*}

This event had its genesis in a halo CME that started at 05:30 UT on
2001 April 10 (http://cdaw.gsfc.nasa.gov/CME\_list/). The starting
radius was $R_{0} = 2.84 \,R_{\odot}$, and  the lateral (plane-of-sky)
expansion speed was $V_{\rm exp} = 2411\,\,{\rm km\, s^{-1}}$. Based
on an examination of several halo CMEs, Schwenn et al.~(2005) conclude
that the radial speed $V$ is typically 0.88 times the expansion speed
$V_{\rm exp}$ of a halo CME. This gives $ V = 0.88\, V_{\rm exp} =
2121\, {\rm km\,s^{-1}}$. The corresponding magnetic cloud was detected
by the ACE spacecraft at 04:00 UT on 2001 April 12, while the shock
reached the Earth at 15:18 UT on 2001 April 11 (Lynch et al.~2003).
Munakata et al.~(2003, 2005) have modelled the magnetic cloud
associated with this event as a cylinder with a Gaussian cross-sectional
profile for the cosmic-ray density depletion inside it. They analysed
the cosmic-ray directional anisotropy from muon
detector measurements from several stations. Assuming that this
anisotropy arises from a diamagnetic drift due to the interplanetary
magnetic field and the density gradient inside the magnetic cloud,
they have derived best-fit parameters for the 3D cylindrical model of
the magnetic cloud as it passes across the Earth. The length of the
flux rope at time $t$ is
\begin{equation}
L(t) = 2\,\pi\,H(t) = 2\,\pi (V\,t + R_{0}) \, .
\label{eq14}
\end{equation}
The length $L$ at time
$T$ is simply
\begin{equation}
L(T) = 2 \pi \, 1\, {\rm AU} \, . 
\label{eq10}
\end{equation}
It may be noted that we are using the value $R_{0}$ for the initial
value of the observed height $H$ above the solar limb. This is not
quite right for full halo CMEs, since $R_{0}$ is really the observed
lateral extent of the CME. However, this is the only concrete
observational quantity available, and the error is not expected to
contribute significantly to our final result for $\sigma^{2}$
(see \S~7). The time elapsed between the first observation of
the halo CME and the detection of the magnetic cloud by the ACE
spacecraft is $46.5$ hours. The average speed of the magnetic cloud
by the time it reaches the ACE spacecraft is 640.6 ${\rm km\,s^{-1}}$
(Lynch et al.~2003), and the spacecraft is located around
$1.5 \times 10^{6}\,{\rm km}$ from the Earth. We therefore estimate
that the magnetic cloud would have taken $\sim 0.6$ hours to traverse
the distance between ACE and the Earth. The total time
$T = 46.5 + 0.6 = 47.1$
hours. The radius of the magnetic cloud is
$
R(T) = 0.106\,{\rm AU}\, ,
$
and magnetic field at the centre of the cloud is estimated to be
(Lynch et al.~2003)
$
B_{\rm MC} = 17.8\,{\rm nT} \, .
$

\subsection{Forbush decrease}

\begin{figure}
\begin{center}
\includegraphics*[width=0.45\textwidth,angle=0,clip]{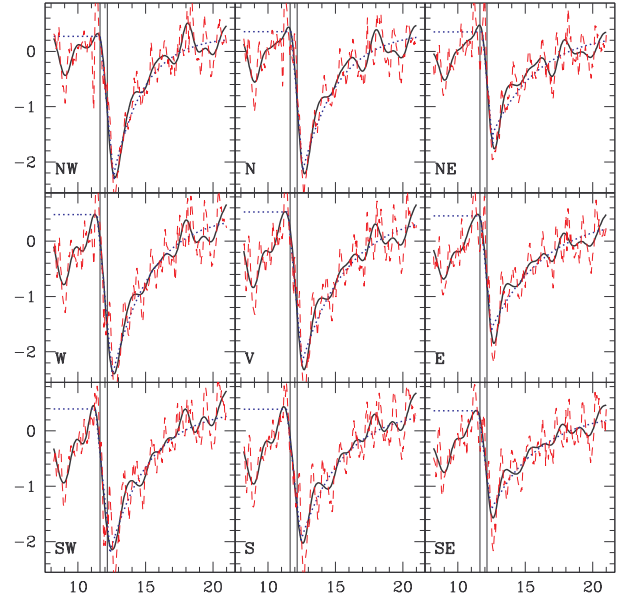}
\vskip -0.1in
\caption{\label{fig010411} The muon flux in the nine directions is
         shown for the Forbush decrease on 2001 April 11. The fluxes
	 are shown as percentage deviation from mean values. The dashed
	 red lines show the unfiltered data, while solid black lines
	 show the data after applying a low-pass filter. The dotted
	 blue lines show the fits to filtered data. The vertical lines
	 in each panel mark the arrival times of shock and magnetic
	 cloud at the Earth.}
\end{center}
\end{figure}

Table 1 gives the start and end times and magnitudes of the Forbush
decrease in different bins for this event, as well as the computed
values of the turbulence level $\sigma^{2}$ (Eq~\ref{eq19}). The shock
associated with this event impacted the Earth at 15:18 UT, 2001 April
11 (Huttunen et al.~2005) and the magnetic cloud first intercepted
the ACE spacecraft at 04:00 UT, 2001 April 12 (Lynch et al.~2003).
Evidently, the magnetic cloud was observed on the Earth after the
start of the Forbush decrease and well before its end in all the
directional bins. It therefore stands to reason that it will have
contributed substantially to the overall Forbush decrease. As
discussed in \S~3, we adopt a value of $\alpha = 0.5$, which means
that the magnetic cloud contributes to 50\% of the overall Forbush
decrease.

\section{2003 October 29 event}
\subsection{Magnetic cloud}
\begin{table*}
\caption{Derived parameters for Forbush decrease for 2003 October 29 event}
\begin{tabular}{llllllllll}
\hline
Quantity & NW & N & NE & W & V & E & SW & S & SE\\
\hline
FD magnitude & 6.81 & 6.78 & 5.21 & 7.76 & 7.76 & 5.85 & 6.92 & 6.82 & 5.26 \\
FD onset & 00:58 & 23:46$^*$ & 22:34$^*$ & 00:58 & 00:00 & 23:17$^*$ & 00:53 & 00:14 & 23:31$^*$\\
FD end & 18:00 & 17:16 & 16:19 & 19:12 & 18:14 & 16:19 & 18:43 & 17:16 & 15:07 \\
$\sigma^{2}$ (Kolmogorov) & 2.48 & 2.49 & 2.55 & 2.45 & 2.45 & 2.52 & 2.48 & 2.49 & 2.55\\
\hline
\noalign{\smallskip}
\noalign{$^*$ These times refer to 2003 October 28.}
\end{tabular}
\end{table*}

This was an exceptional event associated with an X17 flare close to
the central meridian, together with a very fast halo CME. This full
halo CME was first observed at a starting radius of
$
R_{0} = 5.84\,R_{\odot} 
$
at 11:30 UT on 2003 October 28
(http://cdaw.gsfc.nasa.gov/CME\_list/). As with the other  events, the
quantity $R_{0}$ really represents the first observed lateral extent
of the halo CME. The quantity $L(T)$ can be taken to be the same as
defined in Eq.~(\ref{eq10}). The lateral expansion speed of the CME as
discerned from coronograph data is $ V_{\rm exp} = 2459\,\,{\rm km\,
s^{-1}}\, , $ and the radial expansion speed is related to $V_{\rm exp}$
by $V = 0.88\,V_{\rm exp}$ as before. The length of the flux rope at a
given time $t$ is given as before by Eq.~(\ref{eq14}).

Kuwabara et
al.~(2004) have derived best-fit parameters for the 3D cylindrical model
of the magnetic cloud as it passes across the Earth using the methods
described in Munakata et al.~(2003, 2005). They estimate the radius of
the magnetic cloud to be
$
R(T) = 0.14 \,{\rm AU}\, ,
$
the magnetic field at the centre of the cloud to be 
$
B_{\rm MC} = 44\,{\rm nT}\, , 
$
and the time of closest approach of the cloud axis to be 16:27 UT on
2003 October 29. The ACE observations (Huttunen et al.~2005) reveal
that the magnetic cloud made its entry at 12:00 UT, 2003 October 29.
The time elapsed between the first observation of the halo CME
(11:30 UT, 2003 October 28) and the time it was intercepted at the
ACE spacecraft (12:00 UT, 2003 October 29) is  
$
24.5\,{\rm hours} \, . 
$

We could not find data for the magnetic cloud speed as measured by
the ACE spacecraft. We therefore used the estimate of
$\sim 1400\,{\rm km\,s^{-1}}$ from Kuwabara et al.~(2004) for the
magnetic cloud speed. At this speed, the magnetic cloud would take
only $\sim 0.3$ hours to traverse the distance of
$1.5 \times 10^{6}$ km between the ACE spacecraft and the Earth. The
total time $T$ is therefore $T \simeq 24.8$ hours. It may be noted
that the CME liftoff speeds for this event and one previously
discussed (2001 April 11) are similar. However, the event of 2001 April
11 decelerates significantly en route to the Earth, as a result of
which the total travel time $T$ is almost twice as long as for this
event.

\subsection{Forbush decrease}
\begin{figure}
\begin{center}
\includegraphics*[width=0.45\textwidth,angle=0,clip]{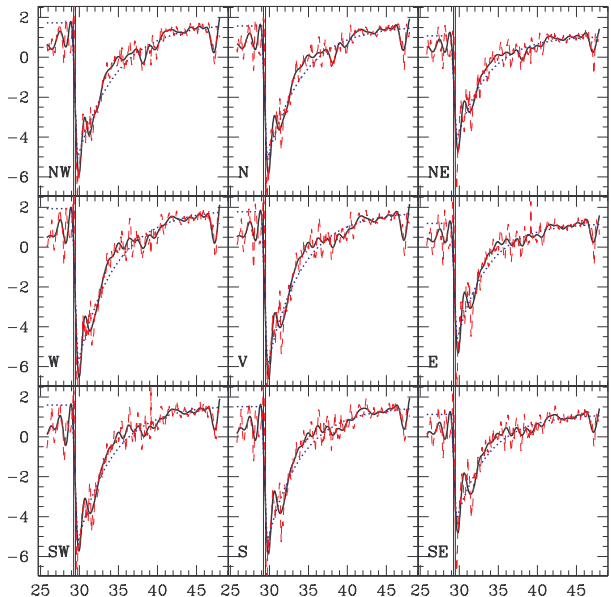}
\vskip -0.1in
\caption{\label{fig031029} The muon flux in each of the nine
         directions for the Forbush decrease on 2003
	 October 29. The line-styles and markings are the
	 same as Fig.~\ref{fig010411}.}
\end{center}
\end{figure}

The Forbush decrease on 2003 October 29 was among the largest observed
by the GRAPES-3 muon telescope, and it is described in detail in
Nonaka et al.~(2006). The time profiles of the decrease as observed by
the different modules is shown in Fig.~10 of that paper. The derived
parameters from the Forbush decrease profiles in the different
directional bins are given in Table 2, where the magnitude of Forbush
decrease derived here from the fit  to the filtered data are
systematically lower than those observed in the raw data, which were
reported earlier (Nonaka et al.~2006). The onset and recovery times in
the 2003 October 29 FD event were relatively faster and thus the
reduction in magnitude is largely caused by the removal of a
higher frequency component due to the use of a low-pass filter to
remove the diurnal variations. The shock ahead of the ejecta impacted
the Earth at 05:58 UT, 2003 October 29 (Dryer et al.~2004), and the
magnetic cloud was detected by the ACE spacecraft at 12:00 UT, 2003
October 29. We noted earlier that it takes around 0.3 hours for the
cloud to travel the distance between ACE and the Earth at a speed of
$1400\,{\rm km\,s^{-1}}$; in addition to this, if one accounts for the
travel time between the leading edge of the cloud and the centre using
the same speed (and a cloud radius of 0.14 AU), it results in an additional
time of 4 hours and 21 minutes. When added to 12:00 UT, the final result
agrees very well with the intercept time of 16:27 UT quoted by Kuwabara
et al.~(2004) for the centre of the cloud. As with
the 2001 April 11 event, the magnetic cloud intercepts the Earth after
the start of the Forbush decrease and well before its end in all the
directional bins. The density depletion inside the magnetic cloud
will therefore account for a substantial part of the overall Forbush
decrease. This justifies our use of  $\alpha = 0.5$ (\S~3) to account
for the fraction of the overall Forbush decrease related to the
CME/magnetic cloud.

\section{2003 November 20 event}
\subsection{Magnetic cloud data}
\begin{table*}
\caption{Derived parameters for Forbush decrease for 2003 November 20 event}
\begin{tabular}{llllllllll}
\hline
Quantity & NW & N & NE & W & V & E & SW & S & SE\\
\hline
FD magnitude & 0.90 & 0.91 & 0.89 & 1.26 & 1.24 & 1.04 & 1.24 & 1.24 & 0.88\\
FD onset & 13:12$^*$ & 11:46$^*$ & 02:10$^*$ & 01:26 & 05:02 & 21:07 & 06:29 & 11:02 & 21:07 \\
FD end$^1$ & 04:57 & 04:26 & 02:32 & 06:00 & 04:54 & 05:00 & 04:16 & 04:12 & 02:54 \\
$\sigma^{2}$ (Kolmogorov) & 0.06 & 0.06 & 0.06 & 0.06 & 0.06 & 0.06 & 0.06 & 0.06 & 0.06\\
\hline
\noalign{\smallskip}
\noalign{$^*$ These times refer to 2003 November 21.}
\noalign{\smallskip}
\noalign{$^1$ The times for FD end refer to 2003 November 24.}
\end{tabular}
\end{table*}

This magnetic cloud event was associated with an erupting filament and
a halo CME, and was observed very well (Wang et al.~2006; Huttunen et
al.~2005) by the WIND and ACE spacecraft. Wang et al.~(2006) have
modelled this cloud as a cylindrical, force-free flux rope to determine
its radius and other parameters. As we shall see, the Forbush decrease
associated with this event is less than those for the other two events
studied in this paper. However, it illustrates the diversity of
events observed with the GRAPES-3 experiment, and is associated with
the best-studied magnetic cloud of the three events considered in this
paper.

The radius of the magnetic cloud as measured by the ACE spacecraft is
(Wang et al.~2006) $ R(T) = T~(7~ {\rm hours}) \times V_{\rm max} =
7 \times 3600\times(750\,{\rm km\,s^{-1}}) = 1.89 \times 10^{7}\, {\rm
km} \, , $ where we have assumed that magnetic cloud moves with the maximum in-situ speed of the ambient solar wind behind the shock $V_{\rm max} = 750$ km s$^{-1}$. The
magnetic field at the centre of the cloud is estimated to be (Wang et
al.~2006) 
$
B_{\rm MC} = 50\,{\rm nT} \, .  
$
The most likely solar origin of this magnetic cloud is a full halo
CME that was first recorded at 08:50 UT on 2003 November 18 at a height
of
$
R_0 = 6.3 R_{\odot} = 2.76 \times 10^{12}\, {\rm cm} 
$
(http://cdaw.gsfc.nasa.gov/CME\_list/). The plane-of-sky expansion
speed $V_{\rm exp}$ of the halo CME as recorded in the CME catalogue
is $V_{\rm exp} = 1660$ km s$^{-1}$. Following Schwenn et al.~(2005),
we computed the radial speed $ V = 0.88\, V_{\rm exp} = 1460
\, {\rm km\,s^{-1}} $. The total time elapsed between 08:50 UT,
2003 November 18 (the time the CME was observed at $6.3 R_{\odot}$)
and 10.1 UT, 2003 November 20 (the time the mag cloud was intercepted
by ACE, Wang et al.~2006) is $1.76 \times 10^5\, {\rm s}$. Since the
ACE satellite is at a distance of $\sim 1.5 \times 10^{6}$ km from the
Earth, we estimate that it takes $\sim 2000$ s for the magnetic cloud
to arrive at the Earth at the solar wind velocity $V_{\rm max} = 750$
km s$^{-1}$. Therefore,
$ T = 1.76 \times 10^5 + 2000 = 1.78 \times 10^{5}\, {\rm s} \, . $

\subsection{Forbush decrease data}

\begin{figure}
\begin{center}
\includegraphics*[width=0.45\textwidth,angle=0,clip]{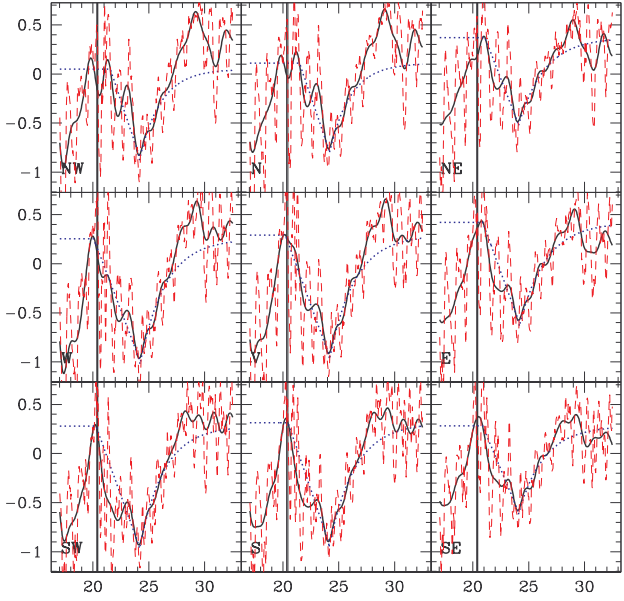}
\vskip -0.1in
\caption{\label{fig031120} The muon flux in each of the nine
         directions is shown for the Forbush decrease on 2003
	 November 20. The linestyles and markings are the
	 same as those for fig.~\ref{fig010411}.}
\end{center}
\end{figure}

Table 3 gives the data derived from the Forbush decrease
observed on 2003 November 20. This is a comparatively smaller event,
and there is considerable variation in start times for the different
bins. This variation is most probably due to uncertainty caused by
oscillations in the fluxes before the decrease (mainly in the NE and
N bins), some of which even remain in the filtered data. In this case
since the magnitude of Forbush decrease is comparatively small, the
diurnal variations are comparable to the decrease, and it is not
possible to fit the unfiltered data to obtain any meaningful
estimates for the parameters of the Forbush decrease; hence, it is
necessary to only use the filtered data for this purpose. Furthermore,
it is very likely that there is some CME-related anisotropy remaining
in the data, and removing it requires the technique used by Munakata
et al.~(2003, 2005) and Kuwabara et al.~(2004), which is outside the
scope of the current paper.

The shock driven by the CME arrived at the ACE spacecraft at 07:27 UT
(Huttunen et al.~2005). Since the average shock speed from the sun to
the Earth is 1020 ${\rm km \, s^{-1}}$ (Wang et al.~2006), it would
take around 1470 s to traverse the $\sim 1.5$ million km between the
ACE satellite and the Earth, and it would have reached the Earth at
around 07:51 UT, 2003 November 20. On the other hand, the ACE
spacecraft first intercepts the magnetic cloud at 10:06 UT (Wang et
al.~2006), and the cloud would have reached the Earth at around 10:36
UT on 2003 November 20, taking some 2000 s to traverse the distance
between ACE and the Earth at the solar wind velocity of $V_{\rm max} =
750\,{\rm km \, s^{-1}}$.  Comparing these times with the start times
of the Forbush decrease given in Table 3, it is evident that the
magnetic cloud start time is before the start time in most of the
bins, and after it in a few of them. However, the end of the Forbush
decrease takes place well after the entry of the magnetic cloud. Thus
the overall decrease should have contributions from both the shock and
the ejecta/magnetic cloud, and we assume that the CME/magnetic
cloud-related component of the decrease accounts for 50\% of the total
decrease, as discussed earlier in \S~3. The estimate of the turbulence
level $\sigma^{2}$ (Eq..~\ref{eq19}) given in Table 3 was computed
according to the procedure discussed in \S~3.

\section{Results} 
We use the observational parameters for each event summarised in
Tables 1--3 using the procedure outlined in \S~3.2 to constrain the
turbulence level $\sigma^{2}$ (Eq.~\ref{eq19}) for each event.  This
quantity represents the ratio of the energy density in the turbulent
magnetic fields to that in the mean magnetic field. We computed
$\sigma^{2}$ separately for each directional bins, and the results are
listed in the last row of Tables 1--3. The average value of
$\sigma^{2}$ for each event is quoted in Table 4.

\begin{table*}
\caption[]{Summary of various observationally derived parameters for
the three events we consider, leading to $\sigma^{2}$ (Eq.~\ref{eq19})
as our final result.}
\begin{tabular}{llllllc}
\hline
Event & $R_{0}$ & $V_{\rm exp}$ & $T$ & $R(T)$ & $B_{\rm MC}$ & 
Average $\sigma^{2}$ (Kolmogorov)\\
&($R_\odot$)& (km s$^{-1}$)& (hr) & (AU) & (nT) \\
\hline
2001 April 11 & 2.84 & 2411 & 47.1 & 0.10 & 17.8 & 2\% \\
2003 October 29 & 5.84 & 2459 & 24.8 & 0.14 & 44 & 249\%\\
2003 November 20 & 6.3 & 1660 & 49.4 & 0.12 & 50 & 6\%\\
\hline
\end{tabular}
\end{table*}

We note that the turbulence levels near the CME/ejecta front for the
2001 April 11 and 2003 November 20 events are fairly small (around
2\% and 6\% respectively). These numbers
are commensurate with what would be expected in the general quiescent
solar wind at a few tens of $R_{\odot}$ (e.g., Spangler 2002). The
turbulence level near the CME/ejecta front of 2003 October 29, on the
other hand, is significantly higher (around 249\%) and is
representative of fairly strong turbulence. This is not
surprising, since this event was a much stronger one, and was
associated with a significantly faster shock. It is conceivable that
strong turbulence associated with the shock front can affect the
vicinity of the CME/ejecta behind it.

Our final result regarding the turbulence level $\sigma^{2}$ is not
very sensitive to the precise value of $\alpha$ (Eq.~\ref{eq9}). A
change of a factor of 2 in $\alpha$ changes $\sigma^{2}$ by 2 to 4\%. We
also pointed out earlier that the estimate for the initial height of
the full halo CMEs is made from the first observation of the lateral
extent of the CME, $R_{0}$. This is the only concrete observable
quantity for halo CMEs, and we have verified that an error of
$1\,R_{\odot}$ in $R_{0}$ will result in an error of around 2\% in the
estimate of the turbulence level $\sigma^{2}$. The decrease
magnitudes vary somewhat from one module to the next, as noted
earlier. We have verified that these differences in the estimate of
the  Forbush decrease magnitude contribute a scatter of only around
2\% to our final result for the turbulence level $\sigma^{2}$. The
parameters that influence our final result most strongly are the
average propagation speed $V$ and the total time $T$ elapsed between
the time when the CME was first observed and when it was intercepted
as a magnetic cloud by the ACE spacecraft.

\section{Summary}

We have focused on the contribution of the CME/ejecta in producing
Forbush decreases observable from the Earth, using data for three
events observed with the GRAPES-3 tracking muon telescope at Ooty. We
selected events that clearly have their origins in full-halo CMEs
that originate close to the centre of the solar disc. The Forbush
decreases produced by these events will therefore be due to the shock,
as well as to the CME/ejecta. While it is somewhat difficult to precisely
separate out the shock-associated and CME/ejecta-associated
contributions, it is reasonable to assume that the CME/ejecta effects
accounts for 50\% of the overall Forbush decrease. We used CME
measurements from LASCO and magnetic cloud measurements from WIND and
ACE to obtain various geometrical parameters of the CME and its
near-Earth manifestation. The CME is a substantially closed magnetic
structure that starts out with a negligible number of high-energy
protons inside it. We consider the diffusion of high-energy cosmic-ray
protons into the CME as it propagates and expands on its way from the
Sun to the Earth. Since the diffusion primarily takes place across the
magnetic fields that enclose the CME, we use an expression from the
literature for the cross-field diffusion coefficient that is derived
from extensive simulations of cosmic rays propagating through
turbulent magnetic fields. The diffusion coefficient depends upon
several quantities like the ratio of the proton Larmor radius to the
maximum scale length of the turbulence $\rho$ (Eq.~[\ref{eq18}]) and
the level of turbulence $\sigma^{2}$ (Eq.~[\ref{eq19}]). The
turbulence near the CME is expected to result in a magnetic-field line
random walk, leading to scattering of protons and contributing to
their cross-field diffusion.

We constrained the magnitude of $\sigma^{2}$ (which is the ratio
of the energy density in the random magnetic fields to that in the
large-scale magnetic field) in the vicinity of the CME front for each
of the events we analysed. We find that the turbulence level
$\sigma^{2}$ near the CME front for the 2001 April 11 Forbush decrease
event is $\sim 2\%$ and $\sim 6\%$ for the event of 2003 November 20,
while it is $\sim 249\%$ for the much more energetic event of 2003
October 29. These estimates may be regarded as an average over the
journey of the CME from the Sun to the Earth, for we do not take
its possible radial evolution into account. The radial evolution of
$\sigma^{2}$ is unlikely to be appreciable, at least in the outer
corona, as we discuss below. If we take turbulent magnetic field
fluctuations to be representative of those in the electron density
(Spangler 2002), we note that the power-law index for electron density
fluctuations $\Delta N_{e}$ as a function of heliocentric distance
ranges from $-2.2$ (Manoharan 1993) to $-1.7$ (Fallows et
al.~2002). This is fairly close to the power law index of the
background electron density $N_{e}$, which is around $-2$ in the outer
corona (e.g., Leblanc et al.~1998). In other words, the quantity
$\Delta N_{e}/N_{e}$ does not vary appreciably with distance from the
Sun, and we can surmise that $\sigma^{2}$ behaves similarly.

To the best of our knowledge, these are the first quantitative
estimates of the turbulence levels near CME fronts. Manoharan et
al.~(2000) have noted that the turbulent density spectra of the
plasma near CMEs are significantly flatter than what is observed in
the slow and fast solar wind, and are similar to what exists in the
near-Sun solar wind acceleration region. However, they were unable to
comment on the level of the turbulence. Badruddin (2002) suggests
that the level of turbulence near the CME/magnetic cloud might
influence the magnitude of the Forbush decrease. However, he
only considered the local effect of the turbulence, while our
treatment takes into account the cumulative effect of the turbulence
near the CME on the cross-field diffusion of protons into
it. Wibberenz et al.~(1998) quote the results of Vanhoefer (1996) who
makes a simple estimate of the cross-field diffusion coefficient by
calculating the ``filling in'' timescale of the CME by energetic
protons. This method results in a rather low estimate for the
diffusion coefficient, and they speculate that this might be due to
the smooth magnetic fields enclosing the CME. Our treatment considers
this aspect in detail, incorporating the expansion of the CME, as well
as results for the cross-field diffusion coefficient adopted from
extensive numerical simulations.

\begin{acknowledgements}
Part of this work was carried out when Prasad Subramanian was in his previous
position at the Indian Institute of Astrophysics.
We thank D. B. Arjunan, A. Jain, the late S. Karthikeyan, K. Manjunath,
S. Murugapandian, S. D. Morris, B. Rajesh, B. S. Rao, C. Ravindran,
K. C. Ravindran, and R. Sureshkumar for their help in the testing,
installation, and operating the proportional counters and the
associated electronics and during data acquisition. We acknowledge
the administrative services of V. Viswanathan. We thank A. A. Basha,
G. P. Francis, I. M. Haroon, V. Jeyakumar, and K. Ramadass for their
help in the fabrication, assembly, and installation of various
mechanical components and detectors. The Japanese members of the
GRAPES-3 collaboration acknowledge the partial financial support
from the Ministry of Education and Science of the Government of
Japan for the GRAPES-3 experiment. We are grateful to the Tibet
neutron monitor group for the use of their data. We thank the referee,
M. Duldig, for several comments and suggestions that have significantly
improved the content of the paper. This paper is dedicated to the memory
of S. Karthikeyan who passed away recently.

\end{acknowledgements}

\end{document}